\documentclass[journal]{IEEEtran}

\ifCLASSINFOpdf
   \usepackage[pdftex]{graphicx}

\else

\fi

\usepackage{xcolor}

\usepackage{amsmath}

\usepackage[caption=false,font=normalsize,labelfont=sf,textfont=sf]{subfig}

\usepackage{url}

\begin{document}

\def\todo#1{\textbf{#1}}

\title{Characterization of MKIDs for CMB observation at 220 GHz with the South Pole Telescope}

\author{%
K. R. Dibert, P. S. Barry, A. J. Anderson, B. A. Benson, T. Cecil, C. L. Chang, K. N. Fichman, K. Karkare, J. Li, T. Natoli, Z. Pan, M. Rouble, E. Shirokoff, and M. Young,

\textit{on behalf of the South Pole Telescope collaboration}
\thanks{
(\textit{Corresponding author: Karia Dibert)}}
\thanks{K. R. Dibert, K. N. Fichman, K. Karkare, T. Natoli, and E. Shirokoff are with the University of Chicago, 5640 South Ellis Avenue, IL, 60637, USA and Kavli Institute for Cosmological Physics, U. Chicago, 5640 South Ellis Avenue, Chicago, IL, 60637, USA (email: krdibert@uchicago.edu, kfichman@uchicago.edu, kkarkare@kicp.uchicago.edu, tnatoli@uchicago.edu, shiro@uchicago.edu) }%
\thanks{P. S. Barry is with Cardiff University, Cardiff CF10 3AT, UK (email: barryp2@cardiff.ac.uk)}
\thanks{A. J. Anderson, B. Benson and M. Young are with Fermi National Accelerator Laboratory, PO BOX 500, Batavia, IL 60510, the University of Chicago, Chicago, IL, 60637, USA, and Kavli Institute for Cosmological Physics, U. Chicago, 5640 South Ellis Avenue, Chicago, IL, 60637, USA (email: adama@fnal.gov;  bbenson@astro.uchicago.edu, myoung@fnal.gov)}
\thanks{T. Cecil, J. Li, and Z. Pan are with the Argonne National Laboratory, 9700 South Cass Avenue, Lemont, IL, 60439, USA (emails: cecil@anl.gov; juliang.li@anl.gov; panz@anl.gov)}%
\thanks{C. L. Chang is with the Argonne National Laboratory, Argonne, IL 60439 USA, the University of Chicago, 5640 South Ellis Avenue, Chicago, IL, 60637, USA, and Kavli Institute for Cosmological Physics, U. Chicago, 5640 South Ellis Avenue, Chicago, IL, 60637, USA (email: clchang@kicp.uchicago.edu) }%
\thanks{ M. Rouble is with McGill University, 845 Rue Sherbrooke O, Montreal, QC H3A 0G4, Canada (email: maclean.rouble@mcgillcosmology.ca)}
}

\markboth{Journal of \LaTeX\ Class Files,~Vol.~14, No.~8, August~2015}%
{Shell \MakeLowercase{\textit{et al.}}: Bare Demo of IEEEtran.cls for IEEE Journals}

\maketitle

\begin{abstract}
We present an updated design of the 220 GHz microwave kinetic inductance detector (MKID) pixel for SPT-3G+, the next-generation camera for the South Pole Telescope. We show results of the dark testing of a 63-pixel array with mean inductor quality factor $Q_i = 4.8 \times 10^5$, aluminum inductor transition temperature $T_c = 1.19$ K, and kinetic inductance fraction $\alpha_k = 0.32$. We optically characterize both the microstrip-coupled and CPW-coupled resonators, and find both have a spectral response close to prediction with an optical efficiency of $\eta \sim 70\%$.  However, we find slightly lower optical response on the lower edge of the band than predicted, with neighboring dark detectors showing more response in this region, though at level consistent with less than 5\% frequency shift relative to the optical detectors. The detectors show polarized response consistent with expectations, with a cross-polar response of $\sim 10\%$ for both detector orientations.
\end{abstract}

\begin{IEEEkeywords}
MKIDs, CMB, South Pole Telescope
\end{IEEEkeywords}

\IEEEpeerreviewmaketitle

\section{Introduction}

 \IEEEPARstart{M}{icrowave} kinetic inductance detectors (MKIDs) exploit the increased inductance produced by the breaking of Cooper pairs in a strip of superconducting metal under an alternating current. When this superconducting strip is coupled to a capacitor the resulting circuit resonates at a specific frequency. If exposed to a sufficiently energetic photon source the resonant frequency shifts as incident photons break Cooper pairs inside the strip, and this frequency shift can be used to monitor the intensity of incident light \cite{zmuid2012}. A simple design, straightforward fabrication, and natural frequency-domain multiplexability make MKIDs well-suited for high-density telescope arrays, and several such arrays have been deployed in recent years \cite{nika2, baselmans2017, auster2018, Brien_2018} or are planned for upcoming years \cite{ccat19, viera20}.

 The subject of this paper is the development of MKID arrays for a next-generation camera for the South Pole Telescope (SPT). The SPT is a ten-meter mm/sub-mm telescope located at the Earth's South Pole whose primary purpose is to measure the cosmic microwave background (CMB). It currently hosts the SPT-3G camera, composed of 16,000 transition-edge-sensor (TES) bolometers \cite{benson2014, bender2019, Sobrin2022}. 
 At the conclusion of the 3G survey, the camera will be replaced with the new MKID-based SPT-3G+ camera. The SPT-3G+ science goals include new constraints on the history of reionization, including the optical depth, via measurements of the kinematic Sunyaev-Zel'dovich effect; characterizing dusty astrophysical foregrounds, which will help enable the discovery of new high-redshift galaxy clusters, distant dusty star-forming galaxies, and new constraints on Rayleigh scattering at recombination \cite{anderson22}.
 MKIDs offer reduced readout complexity and require fewer readout wires, enabling the increased detector density necessary to achieve these science goals. 
 SPT-3G+ will consist of seven monochroic MKID arrays observing at 220, 285, and 345 GHz, with a total of approximately 35,000 detectors across the entire focal plane \cite{dibert2022, anderson22}. Here we focus on the design and characterization of the 220 GHz SPT-3G+ detectors.
 We present the latest 220 GHz pixel design, highlighting changes since \cite{dibert2022}, as well as a dark characterization of the detectors' microwave properties. We also present an optical characterization of 220 GHz prototype detectors, including spectral band measurements, optical efficiency, and polarization response.

\begin{figure*}
    \centering
    \includegraphics[width=0.38\textwidth]{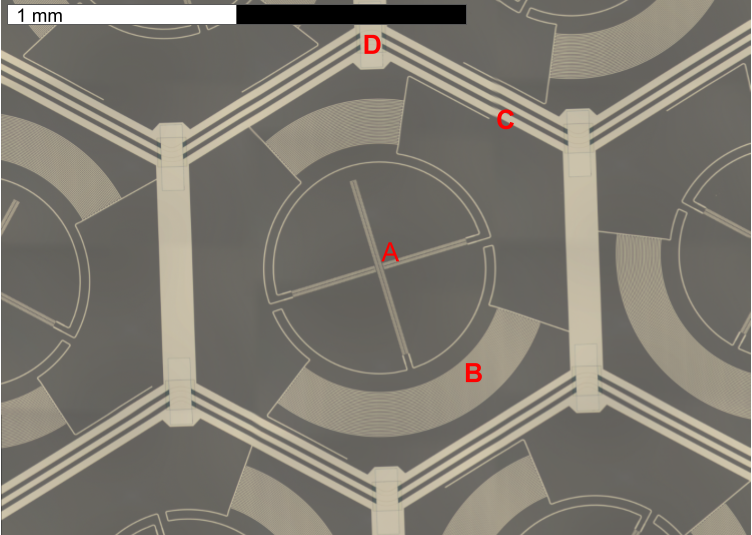}
    \includegraphics[width=0.5\textwidth]{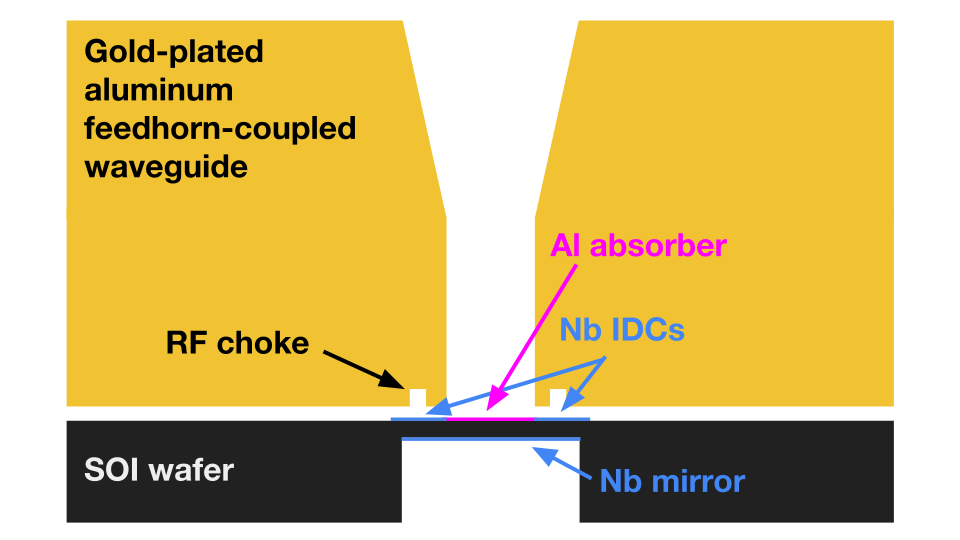}
    \caption{\textit{Left:} Image of an 220 GHz pixel, consisting of two orthogonally aligned inductors (A) coupled to arc-shaped IDCs (B), which are in turn coupled to the coplanar waveguide (CPW) feedline (C). The CPW is bridged at the six vertices of each hexagonal pixel cell (D) to maintain ground-plane connectivity. This device is aluminum-only, but on final devices, structures B, C, and D will be made out of niobium. \textit{Right:} Diagram taken from \cite{dibert2022}, showing the pixel`s position beneath the feedhorn-coupled waveguide and above the integrated silicon quarter-wavelength optical backshort.}
    \label{fig:pixel}
\end{figure*}

\section{Design}

Each SPT-3G+ MKID consists of an aluminum inductor, which acts as a tuned  absorber, coupled to a niobium interdigitated capacitor (IDC). The IDC is then capacitively coupled to a coplanar waveguide (CPW), which serves as a feedline for readout of multiple resonators. A single pixel in the focal plane is made up of two MKIDs whose inductor-absorbers are aligned to orthogonal polarization modes. An image of one SPT-3G+ 220 GHz pixel is shown in the left panel of Figure \ref{fig:pixel}. These sit beneath a feedhorn-coupled waveguide whose diameter sets the lower edge of the observation band; the upper edge of the band is set by a metal-mesh filter in front of the feedhorn. The detectors are patterned onto silicon-on-insulator (SOI) wafers whose device-layer thickness is chosen to act as a quarter-wavelength optical backshort beneath the inductor-absorbers. This integrated silicon backshort allows for post-fabrication editing of the IDCs for fine-tuning resonant frequency placement after testing. The right panel of Figure \ref{fig:pixel} is a cartoon representation of the feedhorn, waveguide, choke, and silicon backshort. Refer to \cite{dibert2022} for details of the inductor geometry and simulated optical bands for each SPT-3G+ observation frequency.

The primary change in pixel design from that published in \cite{dibert2022} is the replacement of the microstrip feedline with a coplanar waveguide. This is due to the strong parasitic coupling between the IDCs and microstrip feedline that appeared in simulations of densely-spaced resonators. To avoid this coupling and maintain control of resonator widths, the CPW, whose field strength declines faster as a function of distance from the feedline, was implemented. The CPW is periodically bridged to maintain ground plane connectivity. A silicon dioxide layer beneath each bridge isolates the feedline in the center of the CPW from the niobium bridge layer.

Fabrication begins with the deposition and patterning of the 120nm niobium IDC and feedline layer, followed by the 30nm aluminum inductor layer. Patches of 1200nm silicon dioxide are patterned via liftoff along the CPW, onto which the niobium CPW bridges are patterned via another liftoff process. Prior to deposition of the bridge layer, the oxide is destroyed using an ion mill at the connection points of the bridges to the ground plane of the CPW. The silicon backshorts are etched from the backside of the wafer down to the device layer, and a layer of niobium is deposited onto the backshort. A thick layer of aluminum coats the rest of the wafer backside, acting as an absorber of stray light.

\section{220 GHz detector characterization}

\begin{figure*}
    \centering
    \includegraphics[width=0.31\textwidth]{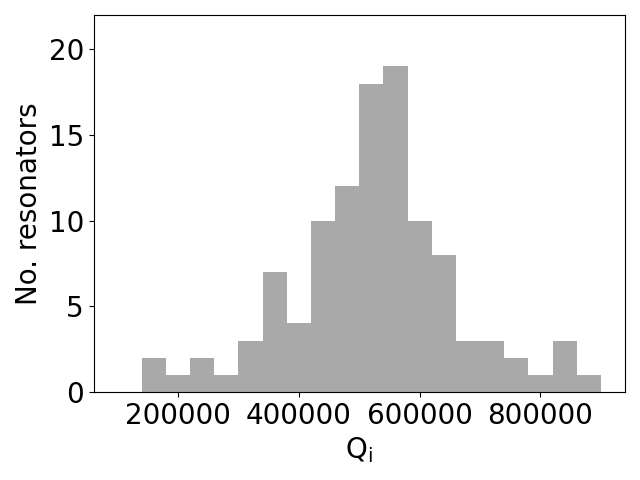}
    \includegraphics[width=0.31\textwidth]{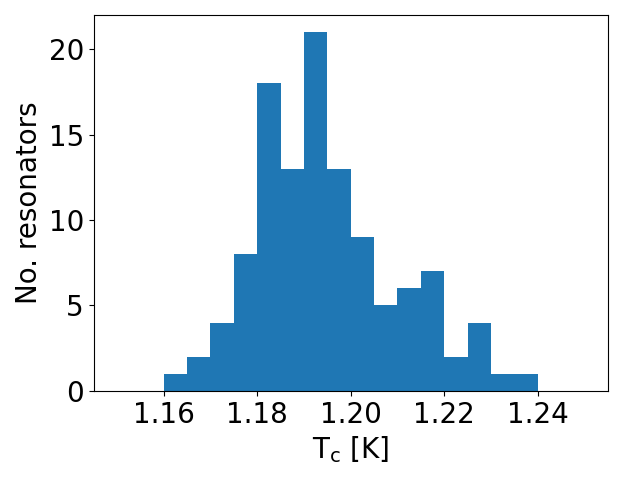}
    \includegraphics[width=0.31\textwidth]{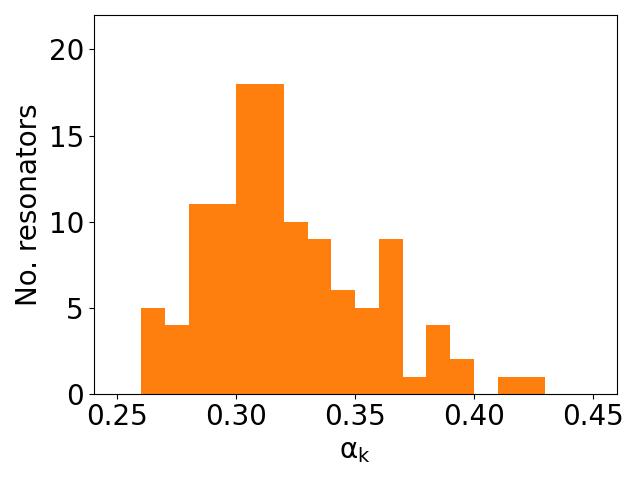}
    \caption{Distributions of inductor quality factor $Q_i$ \textit{(Left)}, Aluminum critical temperature $T_c$ \textit{(Center)}, and kinetic inductance fraction $\alpha_k$ \textit{(Right)} for a 63-pixel (126 resonator) dark 220 GHz array, of which 124 resonators yielded. The $Q_i$ distribution lies above our target multiplexing value of $10^5$. The $T_c$ for the Al film was found to be $\sim$1.2K, and the $\alpha_k$ distribution is consistent with our simulated expectation of 0.34.}
    \label{fig:dark}
\end{figure*}

A detailed characterization of individual dark 220 GHz detectors was presented in \cite{dibert2022}. The findings of that characterization have been consistent across larger dark arrays. Figure \ref{fig:dark} shows histograms of the inductor quality factor $Q_i$ (minimum target value: $10^5$), critical temperature $T_c$, and kinetic inductance fraction $\alpha_k$ (simulated value: 0.34) for a dark chip containing 124 functional dark detectors (97\% yield). $T_c$ and $\alpha_k$ were extracted from fits to the resonant frequency versus temperature curves of the resonators, and $Q_i$ was extracted from fits to the resonance shapes at an operating temperature of 100 mK. The mean $Q_i$ was $\bar{Q}_i = 5\times 10^5$, with $\sigma(Q_i) = 3\times10^5$, with the entire $Q_i$ distribution greater than twice our minimum target value. Assuming a $\Delta = 1.76 k_B T_c$ relationship between the energy gap and critical temperature of the Al film, we found $\bar{T}_c = 1.19$K, $\sigma(T_c) = 0.02$K. The distribution of $\alpha$ was consistent with the simulated expectation, with $\bar{\alpha}_k = 0.32$, $\sigma(\alpha_k) = 0.03$. With dark microwave properties relatively well-understood, the next step on the development timeline is to test the detectors under an optical load.

\subsection{Optical efficiency}

The detectors' response to a controlled optical load was measured in a closed cryostat with the feedhorn-coupled detectors facing a blackbody coldload of variable temperature. Assuming a single-polarization, beam-filling single-moded system, such that the etendue of the system is approximated $A\Omega = \lambda^2 = (c/\nu)^2$, the optical power reaching the detector at a given blackbody temperature $T_{bb}$ can be expressed as the product of the total filter stack transmission $\Gamma_\text{filt}$ and the integral of the blackbody flux:

\begin{equation}
  P_\text{opt}(T_{bb}) = \Gamma_\text{filt} \int_{\nu_{\text{wg}}}^{\nu_\text{filt}}  \frac{ h \nu}{\exp({\frac{h \nu}{k_B T_{bb}})} - 1} d\nu.
  \label{eqn:power}
\end{equation}
The bounds of the integral, $\nu_\text{wg} = 176$ GHz and $\nu_\text{filt} = 240$ GHz, are the lower and upper band edges defined by the waveguide cutoff frequency and metal-mesh low-pass filter frequency respectively. $\Gamma_\text{filt}$ characterizes the band-averaged efficiency of the metal-mesh filters in the cryostat, which is of order 0.70-0.90, depending on the exact optical configuration used. 
Frequency sweeps and timestream data for optically coupled resonators were collected at each blackbody temperature, with the device kept at the intended operating temperature of 140 mK. The frequency vs blackbody temperature curve obtained from the sweeps are converted to detector responsivity $R_x$ via:

\begin{equation}
    R_x (P_\text{opt} ) = \frac{d}{d P_\text{opt}} \Big( \frac{\Delta f_0(P_\text{opt}(T_{bb}))}{f_0} \Big).
    \label{eqn:responsivity}
\end{equation}
The power spectral densities of the  timestream data are fit with a model to extract the detector white noise level $S_{xx}$ at each optical power. Combining the white noise with the responsivity produces an NEP at each optical loading:

\begin{figure}
    \centering
    \includegraphics[width=0.93\columnwidth]{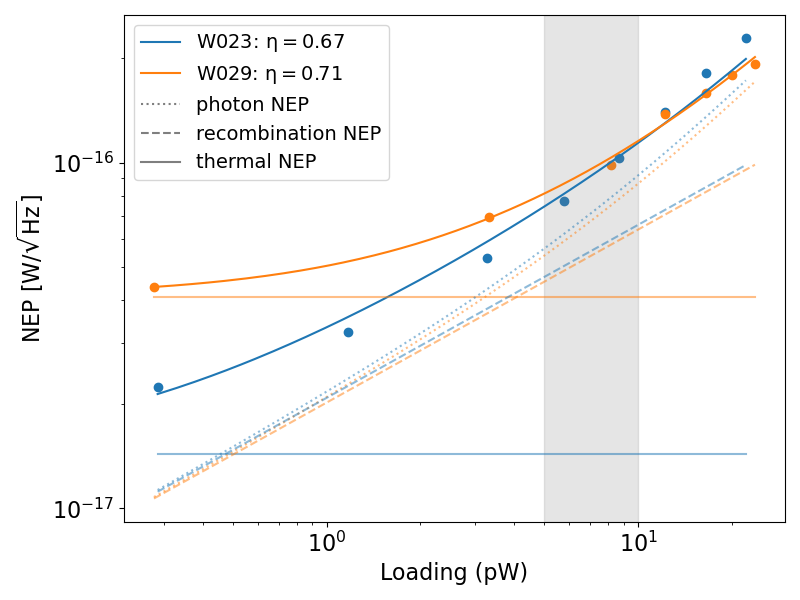}
    \caption{The measured NEP at several different optical loads for both the microstrip (W023, blue) and CPW (W029, orange) chips. These points are fit with a theoretical model (Equation \ref{eqn:opteff}) to extract an optical efficiency, $\eta$, reported in the legend. Fits to the data are represented by solid lines, while other line styles represent the photon, recombination, and thermal components in Equation \ref{eqn:opteff} that make up the total NEP. The gray vertical band is the expected range of optical loading for SPT-3G+.}
    \label{fig:opteff}
\end{figure}

\begin{equation}
    \text{NEP}(P_\text{opt}) = \frac{\sqrt{S_{xx}(P_\text{opt})}}{R(P_\text{opt})}. 
\end{equation}
Two chips were tested with this setup: W023 has the old microstrip feedline, and W029 has the new CPW feedline. Both chips are single-layer aluminum test devices, as opposed to the fiducial design with aluminum inductors and niobium IDCs, feedline and ground plane. (We have not observed a difference in dark responsivity, noise, or resonator quality, between the all-aluminum test devices and the fiducal aluminum and niobium design.) The blue and orange points in Figure \ref{fig:opteff} show the calculated NEP as a function of optical power for both a microstrip (blue) and a CPW (orange) feedline. Within the expected SPT-3G+ optical loading range (gray vertical bar), both resonators have an NEP of roughly $10^{-16}$ $\text{W}/\sqrt{\text{Hz}}$. The loaded NEP as a function of optical power is then fit with a theoretical model \cite{hubmayer2015}:

\begin{equation}
    \text{NEP}^2 = \frac{\text{NEP}^2_\text{photon} + \text{NEP}^2_\text{R}}{\eta_{\text{opt}}} + \text{NEP}_{\text{therm}}^2,
\end{equation}

where $\text{NEP}_{\text{R}}$ is the optical recombination noise, and $\text{NEP}_{\text{therm}}$ is a catch-all for any thermal-like contributions that remain constant with optical power. The photon and optical recombination contributions to the overall NEP are given by:

\begin{equation}
    \begin{split}
    \text{NEP}^2_\text{photon} &= 2 P_{\text{opt}} h \nu \big( 1 + \eta_{\text{opt}} \bar{n}_{\text{ph}}(\nu, T_{\text{bb}}) \big), \\
    \text{NEP}^2_\text{R} &=  2 P_\text{opt} h \nu,
    \end{split}
    \label{eqn:opteff}
\end{equation}

\noindent where $\nu$ is the observation frequency, set to 220 GHz, and  $\bar{n}_{\text{ph}}(\nu, T_{\text{bb}})$ is the mean photon occupation number per mode at the observation frequency for a given blackbody temperature. The optical efficiency $\eta_\text{opt}$ and the thermal term $\text{NEP}_\text{therm}$ are the fit parameters.  The fit to this model is shown in Figure \ref{fig:opteff} for two resonators, a breakdown into the photon, recombination, and thermal NEP is overplotted in dotted, dashed, and solid lines, respectively. Both fits indicate an optical efficiency of $\sim 70\%$. Though this is approximately 15\% lower than the simulated expectation, the simulation does not account for feedhorn loss or other sources of optical power loss.

\begin{figure*}
    \centering
    \includegraphics[width=0.43\textwidth]{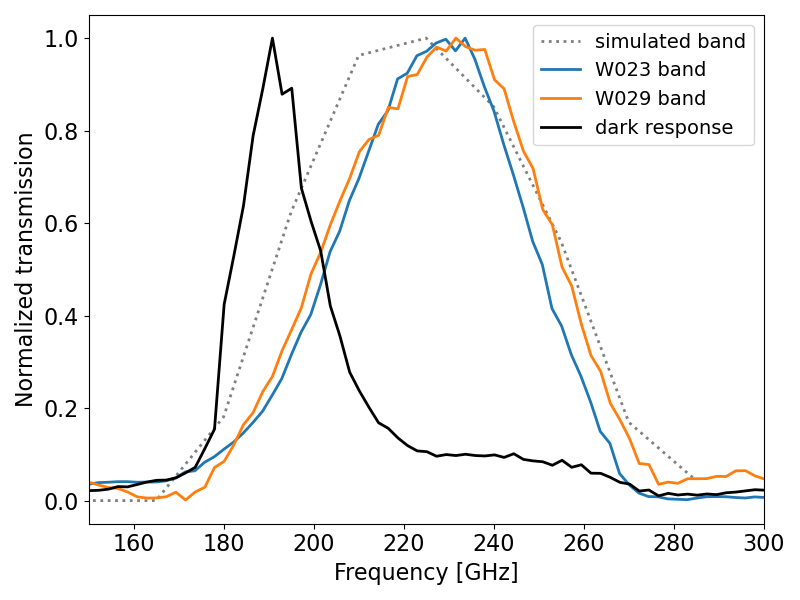}
    \includegraphics[width=0.43\textwidth]{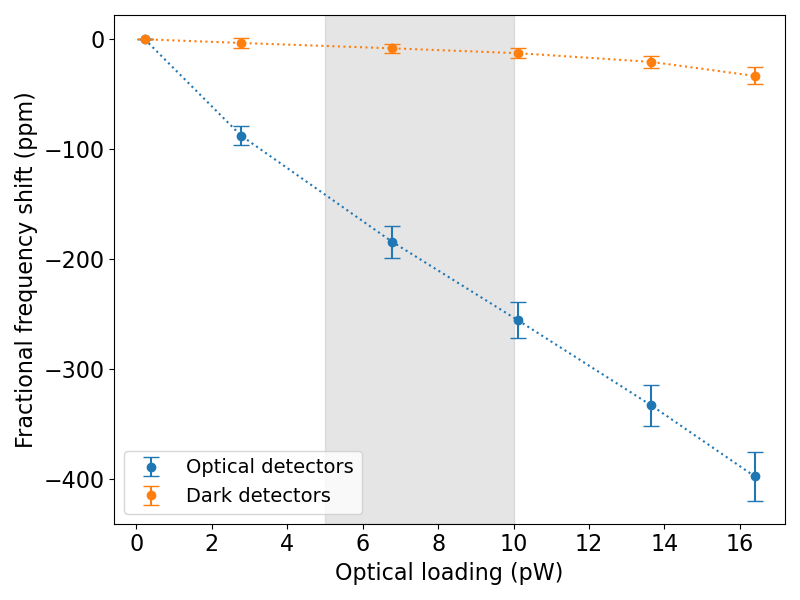}
    \caption{\textit{Left:} The simulated (grey dotted) optical band overplotted with the measured optical band for a resonator from each of the microstrip (W023, blue) and CPW (W029, orange) chips. The black line is the measured spectrum from the dark detectors on W029. All bands here are normalized to have a maximum transmission of 1. \textit{Right:} Comparison of the averaged fractional frequency shift of dark detectors on W029 to that of optically loaded detectors as a function of incident optical power. This ratio of dark to optically loaded frequency shift is less than 5\% within the expected SPT-3G+ optical loading range (gray vertical band).}
    \label{fig:bands}
\end{figure*}

\subsection{Spectral response}
For characterization of the spectral response,
the same devices were tested in a second configuration, this time with the detectors facing an open cryostat vacuum window. The left panel of Figure \ref{fig:bands} shows the spectral response of optical pixels on both the microstrip (W023, blue) and CPW (W029, orange) chips. All spectra on this plot are normalized to have unit maximum transmission. These are overplotted with the simulated optical band, indicated by the gray dotted line. Test results mostly agree with the simulation, except for a discrepancy in the lower part of the band. This could potentially be explained by a slight deviation from simulation in the machined RF choke, or unexpected coupling to the RF choke by nearby structures.

The W029 chip contains several structures in which a feedhorn-coupled central pixel is surrounded by six ``dark'' pixels that are not coupled to a feedhorn. This enables measurements of a dark pixel's response to a change in the power incident on its optically coupled neighbor. The black solid line in the left panel of Figure \ref{fig:bands} shows the measured spectrum of the dark resonators on W029. The dark spectral response peaks at the lower edge of the simulated band, indicating that some of the in-band photons that were not absorbed by the optical detector were scattered and eventually absorbed by neighboring dark detectors. 
The right panel of Figure \ref{fig:bands} shows the averaged fractional frequency shift of dark detectors on W029 as compared to that of optically loaded detectors as a function of incident optical power. Within the expected range of optical loading for SPT-3G+, the ratio of dark to optically loaded frequeny shift is less than $5\%$.  Note that since optical load decreases detector responsivity, this ratio is an overestimate of the expected cross-coupling between two adjacent optically loaded detectors.

\begin{figure}
    \centering
    \includegraphics[width=0.93\columnwidth]{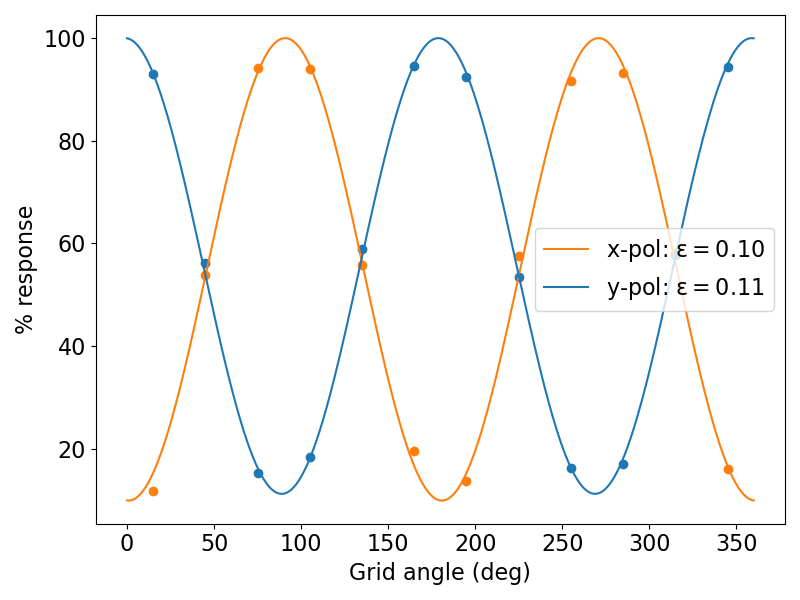}
    \caption{Frequency shift as a percentage of maximum frequency response for two orthogonally-polarized, optically-coupled detectors from W029. The blue and orange points are measurements of fractional frequency shift at various rotation angles of a polarizing grid. Solid colored lines are fits of Equation \ref{eqn:pol} to these points, and the minima of these curves correspond to the percentage of cross-polarization response seen in each detector. This was found to be roughly 10\% for both detectors.
    \vspace{-10pt}}
    \label{fig:pol}
\end{figure}

\subsection{Polarization response}
The polarization response of two resonators with orthogonally-aligned optically-coupled absorbers is shown in Figure \ref{fig:pol}. These measurements were obtained by directing a hot blackbody source onto the detectors through the cryostat vacuum window, with a rotating polarizing grid between the source and detectors. For each polarization angle, the detectors' resonant frequencies were measured with and without the source. The fractional shift in resonant frequency as a function of source polarization angle is fit with the model \cite{ade2015}:

\begin{equation}
    \frac{\Delta f_0}{f_0} (\theta) = A \Big[ \cos(2(\theta - \phi)) + \frac{1+\epsilon}{1-\epsilon} \Big].
    \label{eqn:pol}
\end{equation}

These fits are represented by the solid lines in Figure \ref{fig:pol}, and are rescaled to show frequency shift as a percentage of the maximum response. Here $\epsilon$ is the `cross-polarization parameter,' which corresponds to the minimum of each curve in Figure \ref{fig:pol}. The other fit parameters, $A$ and $\phi$ are an amplitude and phase shift respectively. For both detectors, the cross-polarization was found to be about 10\%. This is higher than the simulated value of 3\%, however, the simulation includes only the waveguide, RF choke, and detectors. The simple conical feedhorn used in this setup is expected to contribute about 4\% to the cross-polarization \cite{goldsmith98}. Further contributions may come from errors in manually-adjusted polarization grid angle, or from stray reflections in the fully on-axis measurement setup.

\section{Conclusion}

Since the publication of \cite{dibert2022}, we have modified the SPT-3G+ pixel design by replacing the microstrip feedline with a coplanar waveguide. Dark testing of a 63-pixel array of these CPW-coupled resonators produces results consistent with those of the microstrip-coupled detectors presented in \cite{dibert2022}. Inductor quality factors are well above our $Q_i = 10^5$ multiplexing target, and the transition temperature $T_c$ and kinetic inductance fraction $\alpha_k$ are tightly distributed around expected values that meet our requirements. Testing of optically-coupled pixels from both the microstrip and CPW feedline devices show good agreement in both the measured optical band and optical efficiency, with both detectors having effectively identical bands and an optical efficiency of $\sim 70 \%$. The measured bands agree with predictions from simulations, except at the lower band edge, where the detector response is less than expected. The polarization of the CPW-coupled optical pixels was also measured, with cross-polarization response of $\sim 10 \%$ for both polarities. Dark detectors on the CPW chip were found to have less than 5\% of the frequency shift of their optically loaded neighbors within the 5-10 pW range of optical loading. The majority of the dark spectral response was seen to come from the lower edge of the band, just past the waveguide cutoff. Further simulations are underway to understand and mitigate this effect, potentially by making slight adjustments to the integrating cavity design or by adding dedicated structures to the pixel to absorb stray photons.

\section*{Acknowledgments}

This material is based upon work supported by the National Science Foundation under NSF-1852617 and NSF-2117894. KD is supported by the DOE Graduate Instrumentation Research Award. This work made use of the Pritzker Nanofabrication Facility of the Institute for Molecular Engineering at the University of Chicago, which receives support from Soft and Hybrid Nanotechnology Experimental (SHyNE) Resource (NSF ECCS-2025633), a node of the National Science Foundation’s National Nanotechnology Coordinated Infrastructure. Work at Argonne National Laboratory was supported by the U.S. Department of Energy (DOE), Office of Science, Office of High Energy Physics, under contract DE-AC02-06CH1137. This work was supported by Fermilab under award LDRD-2021-048 and by the National Science Foundation under award AST-2108763. The McGill authors acknowledge funding from the Natural Sciences and Engineering Research Council of Canada and Canadian Institute for Advanced Research.

\bibliographystyle{IEEEtran}
\bibliography{main.bib}

\end{document}